\documentclass[12pt]{iopart}

\usepackage{iopams}  
\usepackage{graphicx}
\usepackage{epsfig}

\begin{document}

\title[]{Nonlinearity and optical bistability caused by local field effects in the interaction between a two level atomic system and a metamaterial}
\author{J L Garcia-Pomar, R Beigang, M Rahm}
\address{ Department of Physics and Research Center OPTIMAS, University of Kaiserslautern, D-67663 Kaiserslautern, Germany}
\address{Fraunhofer Institute for Physical Measurement Techniques IPM, D-79110 Freiburg, Germany}
\ead{garciapomar@physik.uni-kl.de}

\begin{abstract}
We present a theoretical analysis of the nonlinear response of a metamaterial consisting of an array of split ring resonators coupled to a two level atomic system (2LS) without inversion. We calculated the total complex susceptibility and the linear, third and fifth-order nonlinear solution of the Maxwell Bloch equations that describe the coupled system. We analyzed the impact of the nonlinear effects on the dynamics of an active metamaterial/2LS compound in dependence of the magnitude of the external electric field when the 2LS is pumped by an external pump source and population inversion is achieved. We observed that nonlinear damping in the active metamaterial/2LS compound significantly reduces the effect of spiking. Finally, we investigated the conditions under which the system can exhibit the phenomenon of optical bistability. We observed that optical bistability can only be obtained when the atomic volume density of the 2LS attains unphysically high values and thus is not expected in the system under investigation.

\end{abstract}
\maketitle
\section{Introduction}
In the last decade, the research field of metamaterials has been strongly driven by the desire to control or manipulate electromagnetic waves and to realize optical media with electromagnetic properties that cannot be found in conventional or natural materials. In the realms of the rapidly developing and highly multidisciplinary field of metamaterials, several key research directions emerged including negative index materials \cite{Veselago,Pendry1,GNV1}, hyperlensing \cite{Liu}, cloaks and transformation optics \cite{cloak, Rahm}, chiral metamaterials \cite{Zheludevchiral,Wegenerchiral}, among others. To date, most known metamaterials are composed of metallic structures that are embedded in a dielectric background medium. Since metals suffer from significant losses at higher frequencies -- as e.\ g.\ at infrared or optical frequencies -- many successful demonstrations of metamaterial applications in the microwave range don't immediately translate to other photonic frequency bands. In fact, the search for potential applications of the metamaterial technology at optical frequencies is severely obstructed by the aforementioned limitations that are set by the intrinsic loss of the metallic compositions. In 2007, Boardman et al.\ \cite{boardman} proposed to offset the loss in metamaterials by incorporating active (gain) media into metamaterials. Although this idea is simple at a first sight, it turned out that the coupling mechanisms between gain media and metamaterials are complicated and have to be understood in detail to accomplish the desired effect. In the last two years, different groups contributed to the theoretical description of metamaterials with gain \cite{Wegener2008,Stockmanspaser}. For example, Fang et al.\ \cite{Fang} described a self-consistent approach to describe the interaction between a four-level atomic system and a metamaterial consisting of square split ring resonators. Recently, Xiao et al.\ \cite{Xiao} experimentally demonstrated gain in a negative index metamaterial by engineering dye molecules as a gain material into a fishnet structure.

Another intriguing feature of metamaterials at high power levels is the appearance of nonlinear effects at high external fields. The possibility to control the effective parameters of a metamaterial by exploiting nonlinearities was first suggested in Refs.\ \cite{Zharov2003, Lapine2003}. In the work described in \cite{Zharov2003} a nonlinear response of the metamaterial was achieved by incorporating a dielectric with intrinsic nonlinearity into the metamaterial structure. It was found that the nonlinear effect of the composed system can exceed the intrinsic nonlinearity of the dielectric. The reason for this behavior can be explained in terms of a strong enhancement of the local fields in the vicinity of the metallic structures which causes a strong coupling between the nonlinear dielectric and the metamaterial. Consequently, the nonlinear response of the combined system can be much larger than the intrinsic nonlinearity of the dielectric. Nonlinear metamaterials were successfully demonstrated in the microwave regime by introducing varactor diodes as nonlinear elements in a metamaterial consisting of split ring resonators. It was shown that the magnetic resonance could be dynamically tuned by varying the power of the incident microwaves due to the nonlinear response of the metamaterial. Recently, seminal work was contributed regarding the extension of the theoretical description of nonlinear metamaterials \cite{Smith2010d} and the validity of the theoretical framework was verified by experimental demonstrations \cite{Smith2010a, Smith2010b, Smith2010c}.

Here we present an analytic model model that describes the coupling between a nonlinear two-level atomic system (2LS) and a split-ring resonator (SRR) metamaterial by means of a density matrix formulation. We show that the coupling between the nonlinear medium and the metamaterial is strongly enhanced by local field effects. We studied the interaction for two different regimes. In the first regime, we investigated the nonlinear effects in a hybrid system consisting of an SRR metamaterial and a passive nonlinear material. We derived a full expression for the complex nonlinear susceptibility describing the steady state of the coupled system and discussed the contributions of the linear and nonlinear susceptibilities of third and fifth order to the total response of the hybrid material. We showed that such a coupled system exhibits optical instabilities under certain operation conditions. In the second regime, we studied the temporal dynamics of the population inversion and the output power of the hybrid system in dependence of the incident electric field when the nonlinear medium provides gain and is actively pumped by an additional pump source. We found that the spiking in the dynamics of the SRR metamaterial coupled to an active gain medium is strongly reduced by nonlinear damping effects.

\section{Basic Model}
\begin{figure}
\begin{center}
  \includegraphics[width=9cm]{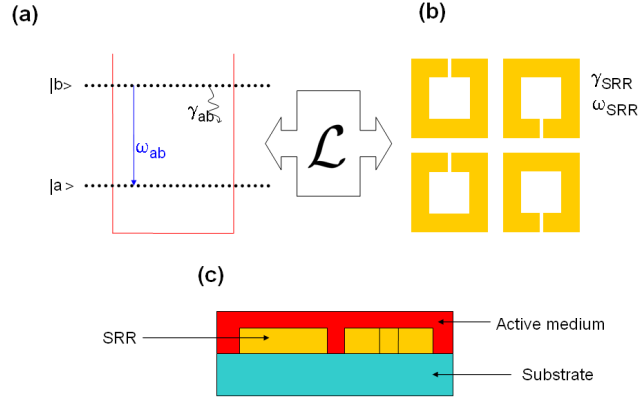}\\
  \caption{Schematic model for (a) a two level system with resonant frequency $\omega_{ab}$ and transversal damping $\gamma_{ab}$ coupled to (b) an array of split ring resonators, with resonant frequency and damping $\omega_{SRR}$ and $\gamma_{SRR}$, respectively, by means of the local fields characterized by the coupling factor $\mathcal L$. (c) Front view of a unit cell of the system where the active medium is surrounding the metamaterial.}\label{geomet}
  \end{center}
\end{figure}

Metamaterials are usually composed of periodic arrays of metallic basis structures with physical dimensions that are significantly smaller than the wavelength of the propagating waves. Dependent on the proper choice of the geometry of the metal structures, metamaterials can provoke a strong local confinement of electromagnetic fields in the medium. For example, the electric field of split ring resonators is concentrated and dramatically enhanced in the gap of the capacitance. In consequence, metamaterials can strongly interact with other media that are placed at locations where the local electromagnetic fields are sufficiently high. It is obvious that the local field enhancement plays a crucial role for the generation of high electric fields to drive the nonlinear polarization in nonlinear media.

In the following, we studied the interaction of a two level atomic system (2LS) with a metallic metamaterial consisting of an array of SRRs. We assumed that the coupling between both systems was caused by the electromagnetic interaction of the local fields that penetrate the SRRs and the atomic two-level medium (Fig. \ref{geomet}). The metamaterial was composed of a single layer of SRRs of copper. The SRRs were described in the dipole approximation whereas the 2LS was treated according to a semi-classical model. The theory of such a model has been thoroughly studied in Ref. \cite{Wegener2008}. However, it should be noted that the model described in \cite{Wegener2008} was inherently limited to linear effects in metamaterials since it only took into account a fixed population inversion in the 2LS that was independent on the magnitude of the incident electromagnetic fields. Therefore, the theoretical approach was not suited for the description of nonlinear effects in metamaterials and did not account for saturation effects in the gain medium. Below we describe a theoretical approach based on a rigorous density matrix description that includes nonlinear effects in a coupled system of a metamaterial and a 2LS and allows to calculate the complex nonlinear susceptibility of the combined system.

To determine the magnitude of the linear and nonlinear response of the hybrid system we described the local fields of the SRRs and the 2LS as a sum of the external field $E_{ext}$ and the local electric fields $E_{L}$. Thereby, we assumed that the local electric fields $E_L^{(1)}$ and $E_L^{(2)}$ at the location of the 2LS and the SRRs were mutually generated by the corresponding polarization fields of the SRRs and the 2LS, respectively. In order to take the coupling efficiency between both systems into account, we introduced a phenomenological coupling constant (local field constant) $\mathcal L$ . By this means, we could describe the local fields as

\begin{eqnarray}
    E_L^{(1)}& = & E_{ext}+\mathcal LP_{SRR}\\
    E_L^{(2)}& = & E_{ext}+\mathcal LP_{ab}
\end{eqnarray}
where $P_{SRR}$  and $P_{ab}$  are the macroscopic polarization of the SRR and the 2LS respectively, given by:

\begin{eqnarray}
  P_{SRR}&=& N_{SRR}\mu_{SRR}\rho_{SRR} \label{eqn:PSRR}\\
  P_{ab} &=& N_{ab}\mu_{ab}\rho_{ab}. \label{eqn:Pab}
\end{eqnarray}
Hereby, $N_{SRR}$  and $N_{ab}$ are the spatial density of SRRs and atoms, $\mu_{SRR}$  and $\mu_{ab}$ are the dipole transition moments of the SRRs  and the 2LS for the transition from the ground state $|a\rangle$ to the excited state $|b\rangle$. Applying the rotating wave approximation (RWA), we expressed the slowly varying amplitude of the coherence $\rho$ as

\begin{equation} \label{RWA}
    \tilde{\rho}=\rho exp(i\omega t)
\end{equation}
and the electric field as
\begin{equation}\label{Eext}
    \tilde{E}_{ext}=E_{ext} \cos(\omega t),
\end{equation}

In this description the diagonal elements denote the probabilities that the SRR or 2LS are in the upper or lower level and the non-diagonal elements determine the coherence between the states $|a\rangle$ and $|b\rangle$ \cite{Boyd, Scully}. From the density matrix formalism we obtained the Maxwell-Bloch equations for the system and could describe the interaction of a 2LS with electromagnetic fields at frequencies that are closely tuned to the resonant frequency of the atomic transition by the level population difference $w$ and the slowly varying off-diagonal matrix element $\rho_{ab}$:
\begin{eqnarray}
    \dot{w} = -\Gamma_{ab}\left(w+1\right)+i\left(\rho_{ab}^{*}\frac{\mu_{ab}}{\hbar}E_L^{(1)}-\rho_{ab}\frac{\mu_{ab}}{\hbar}E_L^{*(1)}\right)+
    \Gamma_{pump}\label{eqn:w2LS}\\
   \dot{\rho}_{ab} = -i\left(\Delta_{ab}-i\gamma_{ab}\right)\rho_{ab}+i\frac{\mu_{ab}}{\hbar}E_L^{(1)}w. \label{eqn:rho2LS}
 \end{eqnarray}
Hereby, $\Delta_{ab}=\omega -\omega_{ab}$ is the detuning,  $\Gamma_{ab}$, $\gamma_{ab}$,  are the longitudinal and transversal damping of the 2LS, respectively and $\Gamma_{pump}$ is the external pumping. In the above equations, the value of the level population difference is $-1$ if all electrons are in the ground state, $0$ for equal population in both levels and $+1$ for total inversion.

On the other hand, we derive the density matrix of the SRRs from the Liouville equation as
\begin{equation}\label{eqn:rhoSRR}
\dot{\rho}_{SRR}=-i\left(\Delta_{SRR}-i\gamma_{SRR}\right)\rho_{SRR}+i\frac{\mu_{SRR}}{\hbar}E_L^{(2)}
\end{equation}
where $\Delta_{SRR}=\omega -\omega_{SRR}$ is the detuning of the external field frequency $\omega$ from the resonant frequency $\omega_{SRR}$ of the SRR and $\gamma_{SRR}$ is the damping constant of the SRR.
Solving the equation system composed by equations (\ref{eqn:w2LS}), (\ref{eqn:rho2LS}) and (\ref{eqn:rhoSRR}) we obtained the solution for the coherences $\rho_{SRR}$, $\rho_{ab}$ and the level population difference $w$. By inserting the solutions of the above system into Eqs.\ (\ref{eqn:PSRR}) and (\ref{eqn:Pab}) we calculated the effective electric susceptibility $\chi$  of the system as a function of the macroscopic polarization fields as:
\begin{equation}\label{eqn:chitotal}
    \chi=V_{ab}\frac{P_{ab}(\rho_{ab},w)}{\epsilon_0 E_{ext}}+V_{SRR}\frac{P_{SRR}(\rho_{SRR},w)}{\epsilon_0 E_{ext}}
\end{equation}
Here, $V_{SRR}$  and $V_{ab}$ are the volume of the 2LS and the SRR divided by the total volume of the unit cell of the complete system, respectively. Note that Eq.(\ref{eqn:chitotal}) determines the total susceptibility of the coupled system including all linear and nonlinear terms that describe the response to external electromagnetic fields.

\section{Nonlinearity}
We first investigated the stationary nonlinear behavior of the coupled metamaterial-2LS compound for the regime where the 2LS was purely passive and not pumped by an additional external pump source $\Gamma_{pump}=0$. To understand the nonlinear physical effects in the system it is illustrative to represent the effective susceptibility as a power series expansion of the external electric field by
\begin{equation}\label{eqn:expan}
    \chi=\chi^{(1)}+\chi^{(3)}|E_{ext}|^2+\chi^{(5)}|E_{ext}|^4 +...
\end{equation}
where we consider only odd orders since the metamaterial structure is centrosymmetric, (as the geometry shown in Fig.\ref{geomet}(b)).
To obtain the values of the nonlinear susceptibility of different order in Eq.\ (\ref{eqn:expan}), we analyzed the nonlinearity of the level population difference $w$ by deriving the  power-series expansion with respect to the relative electric field strength $x=|E_{ext}|^2/|E_s|^2$, where $E_s$ is the analog of the saturation field strength\cite{Boyd}. Under the condition that $\Delta_{SRR}=\Delta_{ab}$, the saturation field $E_s$ is given by

\begin{equation}\label{saturation}
    |E_s|^2=\frac{\hbar^2\Gamma_{ab}(\gamma_{ab}+\gamma_{SRR})}{\mu_{ab}^2}.
\end{equation}
Assuming $x$ to be a small quantity , we performed a Taylor series expansion of $w$ retaining only terms up to the second power. For the resonant frequency $\omega=\omega_{SRR}=\omega_{ab}$ the Taylor expansion takes the form:
\begin{equation}\label{wdesardelcero}
    w=-1+\left(\gamma_{ab}^2+\gamma_{ab}\gamma_{SRR}\right)\frac{\mathcal{L}^2N_{SRR}^2\mu_{SRR}^4/\hbar^2-\gamma_{SRR}^2}{\left(\gamma_{ab}\gamma_{SRR}+ \left(\mathcal{L}\mu_{ab}\mu_{SRR}/\hbar\right)^2 N_{ab}N_{SRR}\right)^2} x+ox^2
\end{equation}
 From equation (\ref{wdesardelcero}) we observe that the main variation in the population inversion $w$ stems from the local field effects that are represented by the local field factor $\mathcal L$.

Taking into account that the solutions of the slowly varying amplitude of the coherence $\rho$, from Eqs. (\ref{eqn:w2LS}), (\ref{eqn:rho2LS}) and (\ref{eqn:rhoSRR}), are given by \cite{Wegener2008}

\begin{equation}\label{rhoab}
    \rho_{ab}=\frac{-w\left(\Omega_{ab}+\frac{\Omega_{ab}\mathcal L N_{SRR}\mu_{SRR}^2}{\hbar(\Delta_{SRR}-i\gamma_{SRR})}\right)}{\Delta_{ab}-i\gamma_{ab}+w\frac{(\mathcal{L}\mu_{ab}\mu_{SRR}/\hbar)^2N_{ab}N_{SRR}}{(\Delta_{SRR}-i\gamma_{SRR})}}
\end{equation}
\begin{equation}\label{rhosrr}
    \rho_{SRR}=\frac{\Omega_{SRR}+\rho_{ab}N_{ab}\mathcal L \mu_{ab} \mu_{SRR}/\hbar}{\Delta_{SRR}-i\gamma_{SRR}}
\end{equation}
where $\Omega_{ab}=\mu_{ab}E_{ext}/\hbar$ and $\Omega_{SRR}=\mu_{SRR}E_{ext}/\hbar$ are the Rabi frequencies for the 2LS and the SRR, respectively, we observe from Eq. (\ref{rhoab}) that the change of $w$ involves two nonlinear effects: the amplitude change of the susceptibility response and the modification of the resonance frequency.

\begin{figure}
\begin{center}
  \includegraphics[width=13cm]{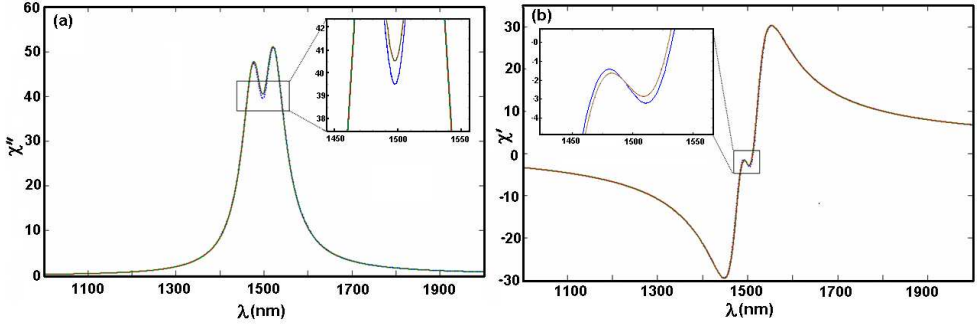}\\
  \caption{Imaginary part (a) and real part (b) of the nonlinear susceptibility at different wavelengths $\lambda$  of the complete system for a high electric external field $|E_{ext}|$. The insets show the differences between the linear (blue dashed line), third order (green dotted line) and complete model (red solid line) with values of $\omega_{SRR}=\omega_{ab}=200GHz$, $\mu_{SRR}=10^{-25}C\cdot m$, $\mu_{ab}=6.5\cdot10^{-29}C\cdot m$, $\gamma_{SRR}=3.4\cdot10^{13}s^{-1}$, $\gamma_{ab}=1.5\cdot10^{13}s^{-1}$, $\Gamma_{ab}=10^{10}s^{-1}$, $N_{SRR}=5\cdot10^{20}m^{-3}$ and $\mathcal{L}=1.7\cdot10^10 m/F$ taken from the simulations and particular values of $N_{ab}=3\cdot10^{23}m^{-3}$ and $x=10^{-5}$}\label{fig2}
  \end{center}
\end{figure}
In a first step, we calculated the total nonlinear susceptibility by the set of equations (\ref{eqn:w2LS}) to (\ref{eqn:chitotal}). As an example, we plotted the imaginary part (Fig. \ref{fig2}(a)) and real part (Fig. \ref{fig2}(b)) of the total nonlinear susceptibility in dependence of the frequency for $x=10^{-5}$ and $N_{ab}=3\cdot 10^{23}$\,m$^{-3}$. We also investigated the contribution of the nonlinear response of different orders to the overall response by inserting the Taylor expansion of Eq. (\ref{wdesardelcero}) into the set of equations (\ref{rhoab}) to (\ref{rhosrr}). As can be seen from Fig. \ref{fig2}, the nonlinear susceptibility reveals a resonant behavior and a pronounced minimum at the resonant frequency. A comparison between the linear (blue dashed line), third order (green dotted line) and total (red solid line) nonlinear response is shown in the insets of Figs. \ref{fig2}(a) and \ref{fig2}(b). From the calculations, we concluded that the deviations between the complete model and the approximated model were maximal at the resonant frequency and therefore in the minimum of the imaginary part of the nonlinear susceptibility. Calculating the imaginary part of the total nonlinear susceptibility at the resonant frequency we obtained
\begin{equation}\label{eqn:chiimag}
    \chi''\approx \frac{N_{SRR}\mu_{SRR}}{\gamma_{SRR}}\left(1-\frac{w(\mathcal L \mu_{ab}\mu_{SRR}/\hbar)^2N_{ab}N_{SRR}}{w(\mathcal L \mu_{ab}\mu_{SRR}/\hbar)^2N_{ab}N_{SRR}-\gamma_{ab}\gamma_{SRR}}\right).
\end{equation}

As can be seen from equation (\ref{eqn:chiimag}), the total nonlinear susceptibility reaches a minimum when the second term in the brackets of (\ref{eqn:chiimag}) approaches to $1$. As a further property, the imaginary part of the total susceptibility and thus the absorption decreases if the local field coupling, represented by the coupling constant $\mathcal {L}$, increases. Furthermore, if the product $\gamma_{ab}\gamma_{SRR}$ is kept constant, the imaginary part of the susceptibility at the resonant frequency decreases for increasing damping $\gamma_{SRR}$ of the oscillations in the plasmonic structure.

\begin{figure}
\begin{center}
  \includegraphics[width=9cm]{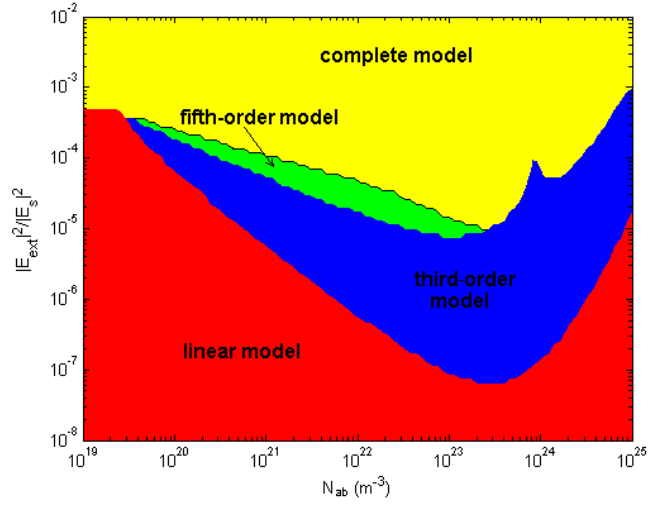}\\
  \caption{Validity ranges for the nonlinear susceptibility of different orders in function of the dipole density $N_{ab}$ of the two level system and the parameter $x=|E_{ext}|^2/|E_s|^2$. The parameters are identical to the parameters chosen in Fig. 2, however $N_{ab}$ and $x$ vary.}\label{fig3}
\end{center}
\end{figure}

In Fig. \ref{fig3}, we identified the validity range of the Taylor expansion in dependence of the atomic density $N_{ab}$ and the relative electric field strength $x$ by comparing the values of the approximated nonlinear susceptibility to the exact solutions of equation system (\ref{eqn:w2LS}) to ({eqn:rhoSRR}). Since we observed that the deviation between the approximated and the complete model were maximal at the resonant frequency of the nonlinear susceptibility as discussed above, we restricted ourselves to calculating and comparing the nonlinear susceptibility at the resonant frequency by applying Eq. (\ref{eqn:chiimag}). As a criterion for the validity of the Taylor expansion we required that the deviation of the imaginary part of the susceptibility that we calculated by inserting the Taylor expansion (\ref{wdesardelcero}) into (\ref{eqn:chiimag}), from the exact solutions should be less than $0.1\%$.

Applying this criterion we could distinguish four different regimes as depicted in Fig. \ref{fig3}. The yellow area represents the range where only the complete model delivers accurate results, i. e. we calculated the total nonlinear susceptibility. As expected, the higher order terms in the Taylor expansion play a more pronounced role when the electric field approaches the saturation limit. Interestingly, the validity ranges are also dependent on the volume density of the SRRs that relates to the mutual interaction between the SRRs and the interaction of the SRRs with the surrounding nonlinear medium. In the green shaded range in Fig. \ref{fig3} we included terms up to the fifth order in the Taylor expansion of the nonlinear susceptibility. As can be seen, the fifth order adds only small corrections to the $\chi^{(3)}$ response of the composite material. Over a wide range of the electric field strength the overall response of the medium can be described by a third order nonlinear process as can be seen from the blue shaded region in Fig. \ref{fig3}. It should be noted that the nonlinearity first increases for higher volume densities of the SRRs up to a maximum value (see the minimum value of the linear response) and then decreases again. This behavior can be seen when the Taylor expansion of the level population difference $w$ in Eq. (\ref{wdesardelcero}) is inserted into (\ref{eqn:chiimag}). For a volume density of the atomic dipoles between values of 10$^{23}$ and 10$^{24}$ we can observe nonlinearities at low field strength. This  is especially interesting as to the observation of nonlinearities in frequency regimes where only low power levels are obtainable as \textit{e. g.} the terahertz frequency range.

\section{Impact of Nonlinear Effects on the temporal Dynamics of an actively pumped metamaterial}
In this section, we analyzed the temporal dynamics of a hybrid material consisting of SRRs and a 2-level atomic system that is actively pumped by external pump fields at a pump rate $\Gamma_0$. To accurately account for saturation effects in the population transfer from the lower to the upper atomic state we described the effective pump rate by $\Gamma_{pump}=(1-w)/2\Gamma_0$. We note that the the effective pump rate $\Gamma_{pump}$ only refers to the process of incoherent pumping, \textit{i. e.} we only pump the populations between the levels $|a\rangle$ and $|b\rangle$, however we do not affect the coherence $\rho_{ab}$ between both states.

\begin{figure}
\begin{center}
\includegraphics[width=12cm]{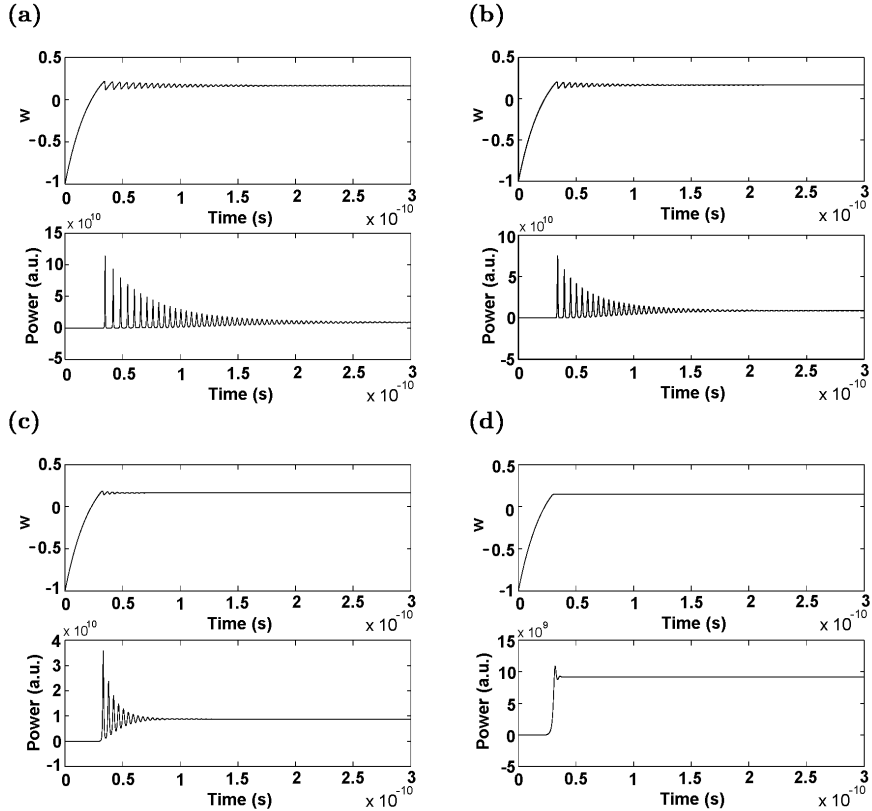}\\
\caption{Temporal dependence of the population inversion $w$ (top) and power of the stimulated emission (bottom) for different external electric fields $E_{ext}$ (a) 10 V/m, (b) 100 V/m, (c) 1000 V/m and (d) 10000 V/m. Same parameters as in Fig. 2 and $\Gamma_{pump}=3.5\Gamma_{ab}.$}\label{dinamica}
\end{center}
\end{figure}

Fig. \ref{dinamica} shows the evolution of the level population difference $w$ and the output power of the hybrid system in dependence of time for different external electric field strengths $E_{ext}$ and a constant pump rate $\Gamma_0=3.5\Gamma_{ab}$. For $E_{ext}=10$ V/m, we observe strong temporal oscillations of the level population difference and the corresponding output power. This typical temporal behavior is well known from laser physics and is refereed to as spiking. The term spiking describes the strong oscillation peaks in the cavity of a continuous wave laser and the emission of strong laser pulses with high electric field amplitudes during the initial turn-on-phase. The effect is caused by the development of a high population inversion far above the laser threshold during external pumping. The absence of appreciable stimulated emission at the beginning of the pump process allows the laser gain to rise substantially beyond the threshold gain. As soon as lasing is initiated, the population inversion is transferred into the stimulated emission of strong laser radiation within very short time, the gain is depleted and the population inversion decreases below the laser threshold. As a result, this single spike in the population inversion corresponds to a first strong pulse that emerges from the laser. Continuous laser pumping allows a periodic revival of population inversion and high laser gain such that we obtain a pulse train of laser pulses with enhanced electric field amplitude. Due to damping in the gain medium and many loss mechanisms in the laser cavity the electric field amplitude of consecutive pulses is damped and after a certain relaxation time the laser process reaches a steady state where the population inversion corresponds to the threshold inversion and the lasing is continuous at constant output power. In the steady state, the emission of laser pulses does not occur and the laser emits a continuous wave.

As can be seen from Figs. \ref{dinamica}(b)-(d), the spiking in the hybrid metamaterial is reduced when the external electric field is increased. At an electric field amplitude of $10000$ V/m the spiking oscillations are almost totally damped out to a single cycle oscillation at the very first start of the amplification process. The damping of the spiking is caused by an increase of nonlinear damping for increasing external electric fields. The nonlinear response in the hybrid metamaterial acts similar to a nonlinear absorber in a laser cavity that reduces the spiking.

\section{Optical bistability}
Optical bistability describes the specific electromagnetic response of a material that supports two (meta-)stable states at a given external field stimulus with an electric field amplitude $E_{ext}$. Optical bistability was first discussed by Hopf, Bowden and Louisell \cite{Hopf}. As meaningful applications, optical bistability can be used for controllable optical switching and the realization of memory cells. In a recent paper \cite{stockman2010}, optical bistability was observed in a spaser emerging from the introduction of nonlinearities by a saturable absorber. In the metamaterial described in this article, we found that optical bistability occurs under certain conditions and can be attributed to local field effects that play a crucial role in the coupling between the SRRs and the two-level atomic medium.

For the investigation of optical bistability we considered the Taylor expansion of the level population difference $w$ that takes the form
\begin{equation}\label{tercerorder}
    w^3+a_2w^2+a_1w+a_0=0
\end{equation}
In the passive case and for $\omega_{ab}=\omega_{SRR}$ and zero detuning $\Delta_{ab,SRR}=0$ between the transition frequency of state $|a\rangle$ and $|b\rangle$ and the SRR resonant frequency $\omega_{SRR}$ we obtained for the coefficients $a_2,a_1$ and $a_0$
\begin{eqnarray}
  a_2 &=& 1+2\frac{A_0}{B_0} \\
  a_1 &=& \frac{A_0}{B_0}\left(2+\frac{A_0}{B_0}+\frac{\gamma_{ab}+x\gamma_{SRR}}{A_0 B_0}\gamma_{ab}\left((\mathcal {L} N_{SRR}\mu_{SRR}^2/\hbar)^2-\gamma^2_{SRR}\right)\right) \\
  a_0 &=& \left(\frac{A_0}{B_0}\right)^2
\end{eqnarray}

where $A_0$ and $B_0$ are described by

\begin{eqnarray}
  A_0 &=& -\gamma_{ab}\gamma_{SRR}\label{Aparameter} \\
  B_0 &=& \left(\mathcal{L}\mu_{ab}\mu_{SRR}/\hbar\right)^2 N_{ab}N_{SRR}\label{Bparameter}
\end{eqnarray}

As can be seen from Eqs. (\ref{Aparameter}) and (\ref{Bparameter}), $A_0$ solely depends on the damping of the atomic transition $\gamma_{ab}$ and the SRR resonance $\gamma_{ab}$ and $B_0$ involves the dipole moments $\mu_{ab}$ and $\mu_{SRR}$ of the atomic transition and the SRR, the volume densities $N_{SRR}$ and $N_{ab}$ of the SRRs and the atoms and especially the local coupling constant $\mathcal {L}$ that represents the local field effects in the system.

As a condition for optical bistability it is required that the three roots of Eq. (\ref{tercerorder}) are real and that at least two of these solutions are stable in the range between -1 and +1 in order to provide physically meaningful multiple solutions. We calculated the roots of ((\ref{tercerorder})) by numerical solution of the equation system (\ref{tercerorder}) to (\ref{Bparameter}).

\begin{figure}
\begin{center}
$\begin{array}{c@{\hspace{0.1cm}}c}
\multicolumn{1}{l}{\mbox{\bf(a)}}&
   \multicolumn{1}{l}{\mbox{\bf (b)}} \\
\epsfig{file=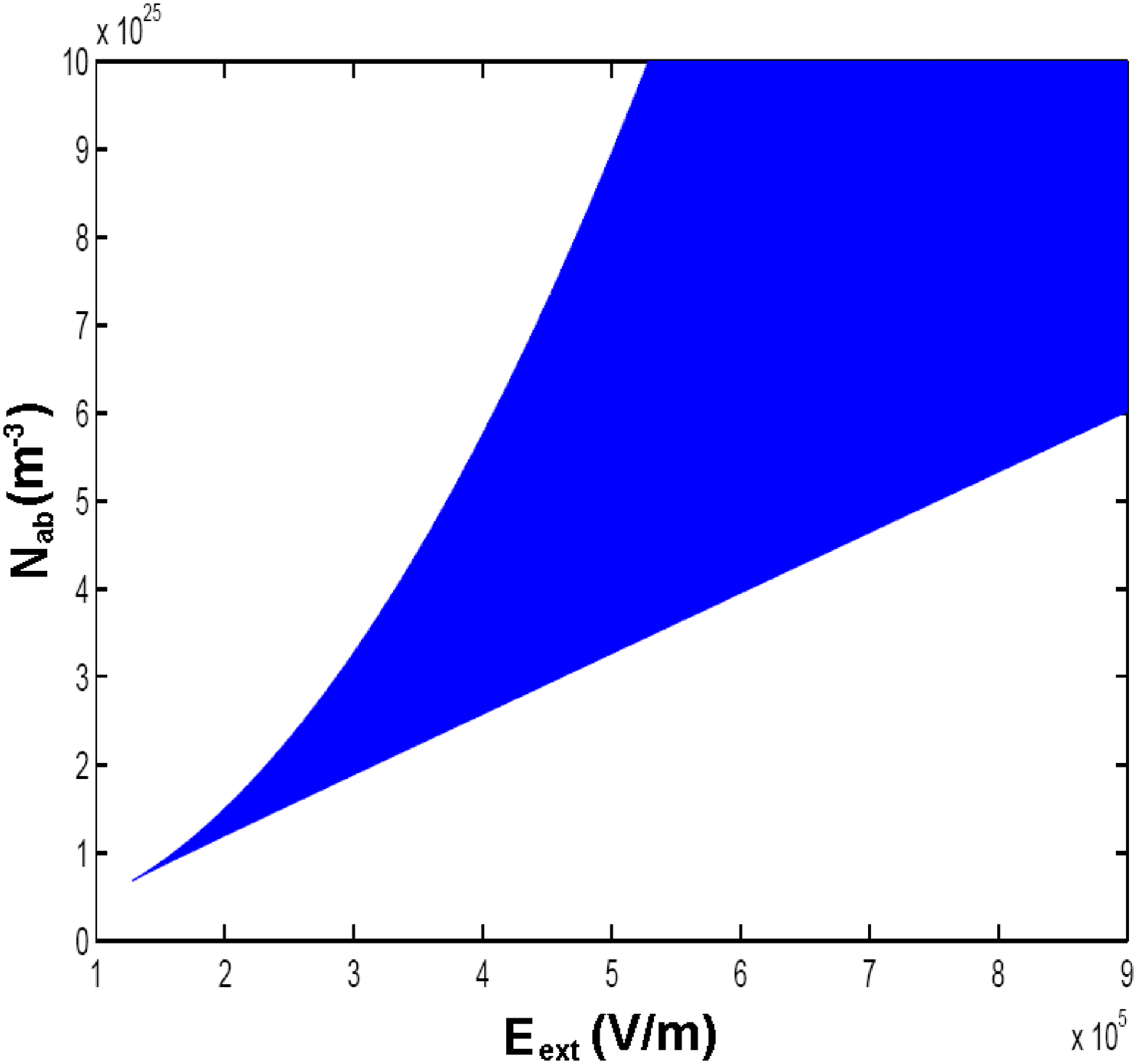,width=8cm, height=6.5cm} & \epsfig{file=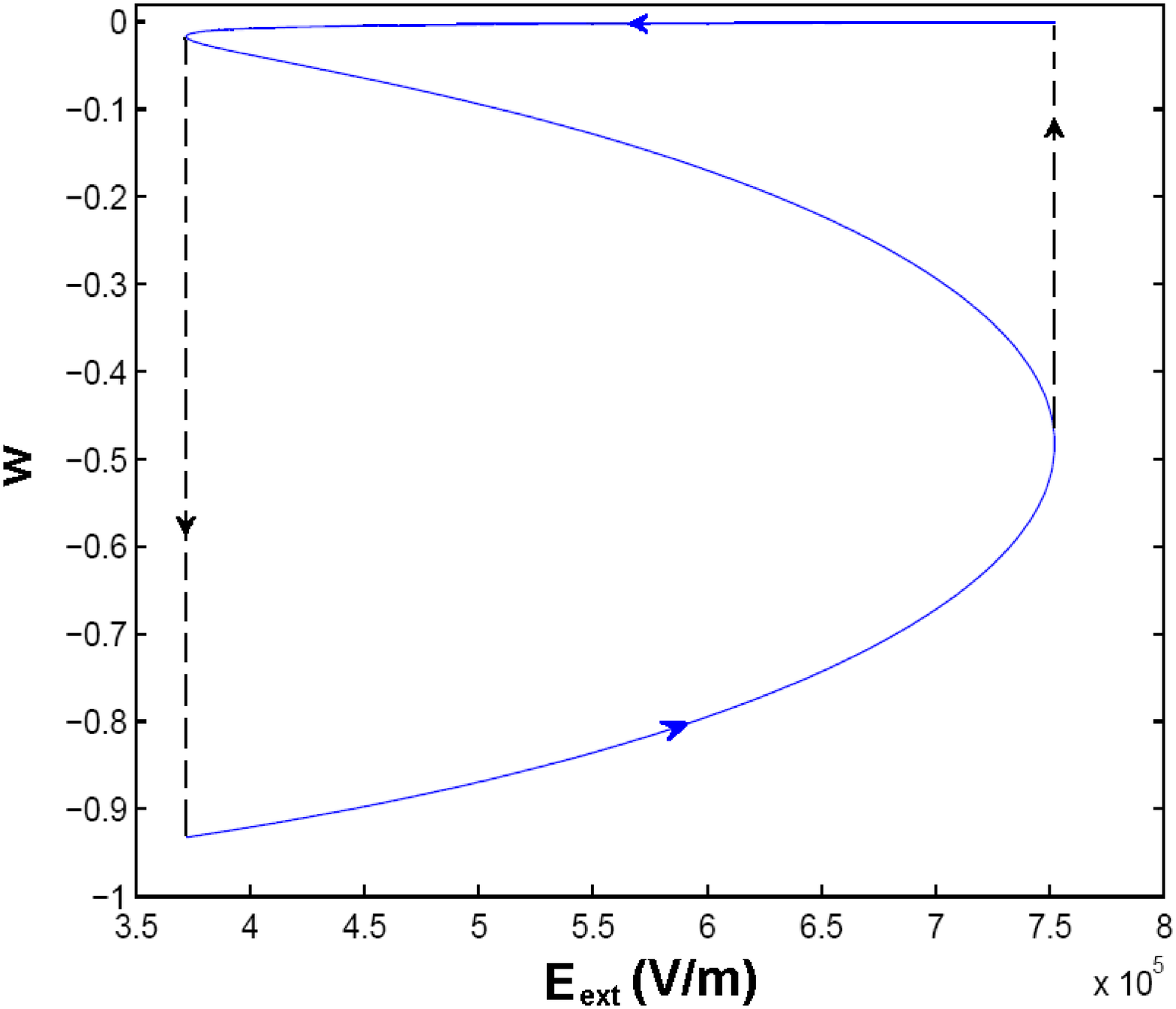,width=8cm,height=6.5cm}\\
\end{array}$
  \caption{(a) Atomic density $N_{ab}$ versus external electric field $E_{ext}$. The blue zone indicates the value range where optical bistability can be observed. Same parameters as in Fig. 3. (b) The three solutions of the population inversion for an atomic density $N_{ab}=5\cdot10^{25}$ $m^{-3}$}\label{bistable}
\end{center}
\end{figure}

In Fig. \ref{bistable}, we show the value range of the external field $E_{ext}$ and the atomic density $N_{ab}$ where the condition for optical bistability is fulfilled. Optical bistability can only be obtained for high external electric fields and extremely high volume densities of the atoms of the gain medium (Fig. \ref{bistable}(a)). The bistable range constantly broadens for increasing electric fields and atomic densities. We note that such high atomic densities cannot be achieved in bulk materials and therefore the investigated system is not suitable for obtaining optical bistability for the investigated range of values.

Fig.\ \ref{bistable}(b) illustrates the level population difference $w$ in dependence of the applied external electric field $E_{ext}$ at $N_{ab}=5\cdot10^{25}$ $m^{-3}$. As indicated in the figure, we can identify two stable branches at a level population difference $w=[0,-0.02]$ that approaches zero and thus the population is equally distributed between the upper and lower atomic level and at $w=[-0.48, -0.93]$ where almost all atoms are in the ground state.

\section{Conclusions}

We investigated the nonlinearities arising from the coupling of an array of split ring resonators (SRR) with a two level atomic medium (2LS) by introducing an analytic model that describes the mutual interaction between the SRRs and the 2LS by local field coupling. Based on this model, we calculated the total nonlinear susceptibility of the system and extracted the linear, third order and fifth order nonlinear contribution by use of a Taylor expansion. We identified the ranges of the nonlinearities of different orders in dependence of the atomic densities of the 2LS and the external electric field. Furthermore, we investigated the temporal dynamics of the coupled system under the presence of external pumping and population inversion in the 2LS. We demonstrated that nonlinear damping significantly reduces the impact of spiking during the built-up-time of the steady state in the system. Finally, we calculated the ranges where optical bistability could occur in the system under investigation. We determined that optical bistability could only be achieved under the condition of unphysically high atomic densities and therefore is not expected to play a role in the investigated system.

\section*{Acknowledgements}
J.L. G-P. acknowledges financial support from Consolider nanolight (CSD2007-00046) and Spanish Ministry of Education by means of "Programa Nacional de Movilidad de Recursos Humanos del Plan Nacional de I-D+i 2008-2011". We thank Peter Weis from University of Kaiserslautern for useful comments.

\section*{References}

\end{document}